# Fabrication, optical characterization and modeling of strained core-shell nanowires.

Z. Zanolli[(1)], L.E. Fröberg, M.T. Björk, M-E Pistol, and L. Samuelson

Solid State Physics/The Nanometer Structure Consortium,

Lund University, Box 118, S-211 00 Lund, Sweden

**Abstract**

Strained nanowires with varying InAs/InP core/shell thicknesses were grown using Chemical Beam Epitaxy. Microphotoluminescence spectroscopy, performed at low temperature, was then used to study the optical properties of single wires. Emission from the InAs core was observed and its dependence on the shell thickness/core diameter ratio was investigated. We found that it is possible to tune the emission energy towards 0.8 eV by controlling this ratio. We have compared the measured energies with calculated energies. Our findings are consistent with the wires having a hexagonal crystal structure.

## 1. Introduction

Core/shell strained nanowires have recently become of interest. One of the leading motivations for their study is the fact that the presence of a shell improves the luminescence efficiency of the core, otherwise (i.e. in uncapped wires) limited by surface states. Moreover, the strain between the two different materials will affect the band structure and, consequently, the emission features of the wires: an understanding of this effect allows a tuning of the wire's emission energy [1].

[1] Zeila Zanolli, Solid State Physics, Lund University, Box 118, S-211 00 Lund, Sweden,
phone: +46 (0)46 2224511, Fax: +46 (0)46 222 3637, e-mail: zeila.zanolli@ftf.lth.se

The materials used were InAs for the core and InP for the shell. The wires are fabricated via Chemical Beam Epitaxy (CBE) because this method allows to achieve the growth of defect free and high quality interfaces even between strongly lattice-mismatched materials.

**2. Growth**

The nanowires were grown by CBE [2]. Metal particles act as seed particles for nanowire growth where the interface between the metal and semiconductor is the growing interface. Thus, the nanowires grow with the metal particle on top and the size of the particle defines the diameter of the wire. This is schematically shown in figs. 1.a and 1.b. In this study, Au aerosol particles with diameters of 20 nm and 40 nm were used as the seed particles. The particles are randomly deposited on an InAs (111)B surface and then the samples are transferred to the high vacuum growth chamber. The growth process starts with deoxidation at 510 ºC under a pressure of 1.5 mbar of thermally cracked TBAs (tertiarybutylarsine). During this process step the Au particles alloy with the In in the substrate and the diameter of the seed particle increases slightly. The temperature is then lowered to 420 ºC and TMIn (trimethylindium, 0.2 mbar) is introduced in the chamber. This defines the start of the InAs core growth which lasts for 60 minutes, to achieve 4 µm long wires, and yields high aspect ratio wires that are homogenous in diameter and length. The crystal structure is mainly wurtzite with thin slices of zincblende inserted randomly along the wire. In order to terminate the axial core growth and allow for the radial shell growth, the temperature is lowered to 350ºC and TMIn is switched to TEIn (triethylindium, 0.2mbar) and TBAs to TBP (tertiarybutylphosphine, 2.5 mbar). The result is schematically shown in fig. 1.c.

## 3. PL measurements

The optical characterization was performed by photoluminescence (PL) measurements on single nanowires.

After the growth, the wires were transferred onto a gold patterned Si/SiO$_2$ substrate which allows easy mapping of the wires. At first, the patterned substrate was investigated by using an optical microscope (100X objective, dark field) to find and map isolated wires suitable for the PL analysis. A further check that the measurements were performed on single nanowires instead of a few of them stacked together was performed by using SEM (Scanning Electron Microscope) after PL. The selected wires were usually separated by more than ~20 μm from other wires and any other particles on the substrate.

For the micro-PL measurements a set-up optimized for detection of radiation in the near IR (0.9 μm – 2.5 μm) was employed. It consists of a liquid N$_2$ cooled camera having a HgCdTe focal plane array sensor, a spectrograph, a long working distance microscope with reflective objective, lenses and mirrors, and a laser (frequency doubled Nd-YAG emitting at 532 nm) for the excitation of the samples. The samples are held in a continuous flow helium cryostat for low temperature (~ 5 K) measurements. To collect the luminescence from an excited wire, the emitted light was sent through the microscope, dispersed by the spectrometer, and detected by the camera.

After the PL experiment, the wires were imaged with SEM (fig. 2) and their actual diameter was then measured. Next we calculated the shell thickness, defined in the inset of fig. 5, using the core diameter estimated with respect to the Au particle size. Those values were used as input for the calculations, in such a way to have a direct comparison between theory and experiment.

**4. Calculations**

In order to determine the bandstructure of the whiskers we first calculated the strain-distribution using continuum elasticity theory [3]. The strain-energy minimization was performed using a finite-difference approximation and we used a grid of 120*120*120 elements. Using the so-obtained strain tensor elements we could then find the local band-gap using linear deformation potential theory. Due to the low bandgap of InAs it was necessary to allow mixing of eight bands. We used parameters from the tabulation by Vurgaftman et al. [4]. A typical calculated band structure is shown in fig. 3 where the core and shell thicknesses were equal. Only the ratio of the core and shell thicknesses affects the result since we do not account for confinement [5]. The calculations were performed assuming zinc-blende structures since there are no parameters available for wurtzite structure and only a small difference between zincblend and wurtzite is expected.

**5. Results and Discussion**

Two sets of nanowires were grown. In one set the shell thickness was varied and in the other set the core thickness was varied.

We have examined the behavior of every wire at different excitation power density finding evidence of state filling at high laser power. In the presence of a cw laser beam the population of the energy states can be controlled modifying the excitation power density. When the latter is increased, states at higher energy will be gradually populated (state filling [6]) and will contribute to the PL emission, since the relaxation to lower energy states is forbidden by the Pauli exclusion principle. The resulting spectra are thus wider and shifted towards higher energy with respect to the ones collected at low excitation power density. This blue shift is demonstrated in the spectra of fig. 4 in the case of InAs 40nm core / InP 10nm shell (fig. 4.a) and a InAs 40nm core / InP 20nm shell (fig. 4.b) wires. We note that the emission from single wires was undetectable for the sample with the thinnest (10nm) shell, so in this case the spectrum is obtained by averaging over several wires. This shows the importance of adding a shell to the core to improve the luminescence efficiency, which is confirmed by the more intense emission from wires with thicker ($\geq$ 20nm) shell.

The same figure illustrates the energy shift due to strain: a thicker shell has the effect of increasing the bandgap of the core material, which results in a blue shift of the emitted energy. A similar behavior was observed in strained SK QDs luminescence of ref. 7.

This trend is more general, as shown in fig. 5 where the measured and the calculated energies for wires having different shell thickness to core diameter ratio are displayed. The energy of investigated samples was tuned from 0.66 eV to 0.8 eV.

From this graph it follows that we can change the emission energy further to the 0.95 eV – 0.8 eV region (i.e. 1.3µm – 1.55µm) by raising the shell/core thickness ratio. This can be done either by increasing the shell thickness at fixed core diameter or by decreasing the core diameter at constant shell thickness. It should be noted that in the latter case the blue energy shift is due not only to the strain between InP and InAs but is also due to quantum confinement for sufficiently small diameter of the core. The measured energies and the calculated ones have the same trend, but the experimental values are usually lower than the theoretical ones by ~160meV. This can be explained by the fact that the grown wires are not purely zinc-blend in crystal structure but mainly wurtzite (as confirmed by TEM investigation). Wurtzite structures typically have larger band-gaps than the corresponding zinc-blende structures.

## 6. Conclusion

In summary, we have demonstrated that the emission energy of strained core/shell whiskers can be tuned by varying the shell thickness to core diameter ratio. Due to the good control over the growth conditions this means that bandstructure engineering is possible with nanowires. All the measured spectra were compared with theoretical calculations, finding a good agreement in the trends and a shift in energies that reveals the wurtzite crystal structure of the wires.

## 7. Acknowledgments

This work was performed within the nanometer consortium in Lund and supported in part by VR, SSF and in part by the European Community's Human Potential Program under contract HPRN-CT-2002-00298 and NOE SANDIE under contract NMP4-CT-2004-500101.

**Figures captions**

**Fig 1** Schematic representation of the growth process: a) deoxidation and Au/In alloy, b) core growth, c) shell growth

**Fig. 2** SEM image of a 20nm InAs core/20nm InP shell nanowire.

**Fig. 3** Calculated energy bands for a wire with core diameter equal to shell thickness.

**Fig. 4** PL spectra for InAs/InP 40nm/10nm (a) and 40nm/20nm (b) as a function of excitation power density. The spectra have been normalized and shifted in intensity for clarity. The state filling at high power and the energy shift due to strain are shown.

**Fig. 5** Energy tuning via shell thickness to core diameter ratio: the emission energy versus the ratio *"shell thickness"* to *"core diameter"* is shown. Inset: schematic representation of a core/shell wire section with d = 40 and t = 30.

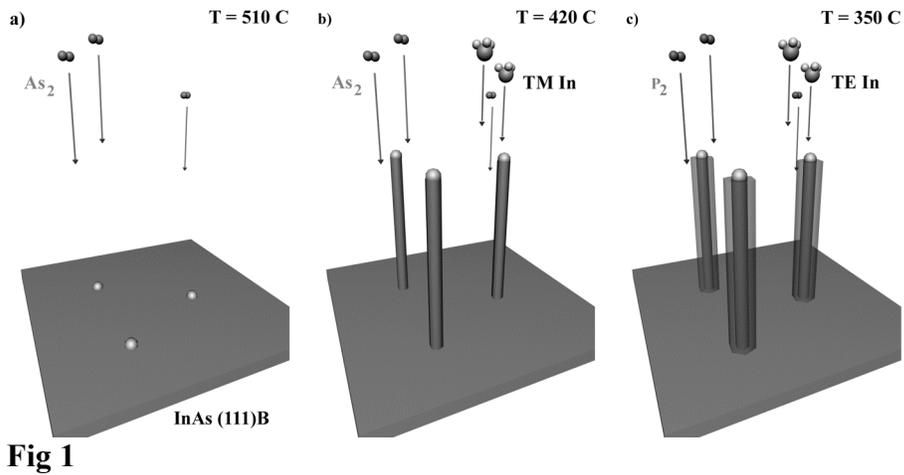

**Fig 1**

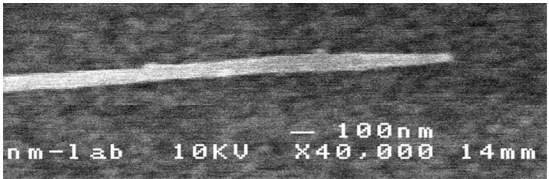

**Fig. 2**

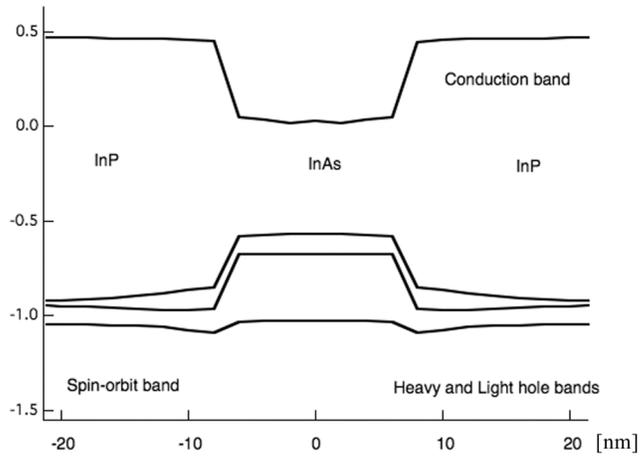

**Fig. 3**

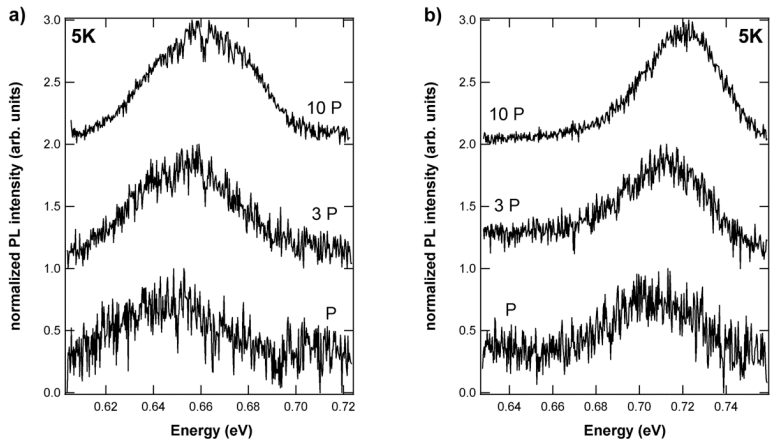

**Fig. 4**

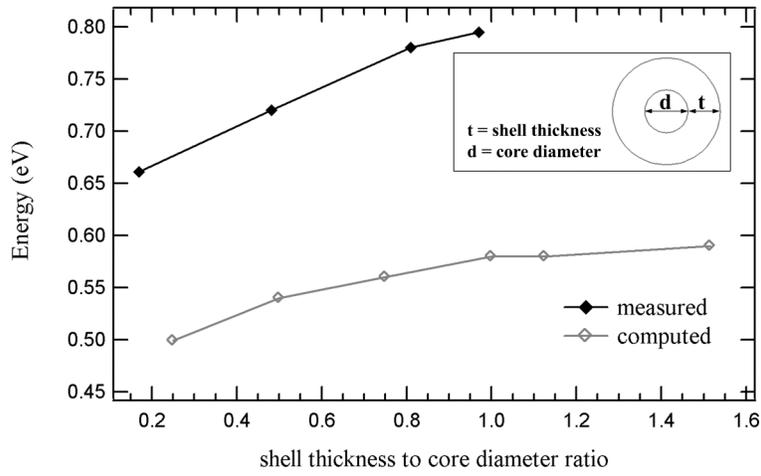

**Fig. 5**

9